\documentclass[a4paper,12pt]{article} 

\usepackage{geometry,epsfig,subfigure}
\usepackage{amsmath}
\usepackage{amssymb}
\usepackage{graphics,epsfig,mydef}
\usepackage[square,sort,comma,numbers]{natbib}

\usepackage[usenames]{color}
\usepackage{fancyhdr,url}   
\usepackage{setspace}  
\usepackage{multirow}   
\usepackage{rotating}    
\usepackage{colortbl}     
\usepackage{relsize} 
\usepackage{booktabs}
\usepackage{cancel}
\usepackage{enumerate}
\numberwithin{equation}{section}
\usepackage{soul} 
\usepackage{xcolor} 
\usepackage{pdflscape}
\usepackage{hyperref}
\usepackage[capitalise]{cleveref}
\usepackage{lineno}
\usepackage{graphicx}
\usepackage{etoolbox}
\usepackage{epstopdf}

\definecolor{darkblue}{rgb}{0,0,0.8}
\definecolor{darkgreen}{rgb}{0,0.5,0}

\long\def\symbolfootnote[#1]#2{\begingroup \def\thefootnote{\fnsymbol{footnote}}\footnote[#1]{#2} \endgroup} 

\renewcommand{\cos}[1]{ \text{cos}\hspace{0.0cm}\left( {#1} \right) }

\usepackage{amsthm}

\newcommand{\HRule}{\rule{0.9\linewidth}{0.2mm}}

\begin{document}
\renewcommand*{\thepage}{\arabic{page}}

\setstretch{1.3}

\begin{center}
\large
\textbf{A First Look at the Performance of Nano-Grooved Heat Pipes\\}

\normalsize
\vspace{0.2cm}
Yigit Akkus$^{a,b}$, Chinh Thanh Nguyen$^{a}$, Alper Tunga Celebi$^{a}$, Ali Beskok$^a\symbolfootnote[1]{e-mail: \texttt{abeskok@smu.edu}}\!$ \\
\smaller
\vspace{0.2cm}
$^a$Lyle School of Engineering, Southern Methodist University, Dallas, Texas 75205, USA \\
$^b$ASELSAN Inc., 06172 Yenimahalle, Ankara, Turkey\\
\vspace{0.2cm}
\end{center}

\begin{center} \noindent \HRule \\ \end{center}
\vspace{-0.6cm}
\begin{abstract}

\noindent Passive thermal spreaders utilizing liquid/vapor phase-change mechanism such as heat pipes, have been widely used in the macro-scale thermal management of electronic devices for many years. Micro-fabrication techniques enabled the fabrication micro-scale grooved heat pipes on semiconductors. Recent advances in fabrication techniques, on the other hand, enabled producing nano- and \aa ngstr\"{o}m-scale capillaries and cavities, which renders the manufacturing of nanoscale heat pipes possible. In the present study, we have simulated nanoscale heat pipes composed of nano-grooves using molecular dynamics and evaluated their performance based on different operating parameters such as the filling ratio and heat load. Moreover, evaluation of size effect on the thermal performance is made by comparing proportionally scaled heat pipes. Simulation results reveal that efficient operation of nano-grooved heat pipes depend not only on the proper selections of filling ratio and heat load, but also on the geometrical parameters such as cross sectional dimensions and aspect ratio of the groove. The modeling strategy used in this study opens an opportunity for computational experimentation of nanoscale heat pipes.

\vspace{0.2cm}

\end{abstract}
\vspace{-0.6cm}
\begin{center} \noindent \HRule \\ \end{center}

\section{Introduction}
\label{sec:intro}

Transistor density on an integrated circuit (IC) has been steadily increasing as suggested by Gordon Moore more than half a century ago \cite{moore1965}. Reduction in the size and increase in the component density lead to enormous heat flux values for today's electronic and photonic devices. For instance, radiofrequency monolithic microwave ICs using GaN high-electron-mobility transistors can generate a heat flux of $1000\unit{kW/cm^2}$
at the gate area \cite{calame2007,bagnall2013}. Thermal management of high heat flux (HHF) devices is vital to realize the proper functioning without deterioration of performance and longevity. It should be stated that, system-level thermal management is not the critical issue due to the availability of capable conventional cooling methods \cite{qu2017}. Chip-level cooling, on the other hand, is the major bottleneck for proper functioning of devices due to the formation of local hot spots with large on-chip temperature gradients \cite{perpina2011,tavakkoli2016}.

Thermal scientists have long been seeking efficient techniques to cool HHF devices. Single-phase cooling methods are impractical due to the high temperature difference and pumping needs \cite{prasher2005}. Phase-change methods are desirable addressing their advantage in high latent heat of evaporation, which enables removal of high amount of heat with small temperature difference. Passive thermal spreaders utilizing liquid/vapor phase-change mechanism such as vapor chambers and heat pipes, are widely used in electronic cooling. Closed loop circulation of the working fluid is provided by the capillary pumping in liquid phase and density gradient in the vapor phase. While the vapor chambers have excellent hot spot removal ability by transporting the localized heat from a source to a large heat rejection surface, heat pipes function as superconductors to remove the waste heat away from the source.

The first heat pipe was using wire mesh wick structure to pump the condensate to the evaporator section \cite{grover1964}. Then a single sealed non-circular micro-channel, whose sharp-angled corners work as liquid arteries, was proposed as micro heat pipes \cite{cotter1984}. To increase the heat carrying capacity of a heat pipe, the number of liquid arteries should be maximized. Consequently, grooved heat pipes, which utilize multiple grooves machined on the inner wall of the base metal as the wick structures, have emerged and been widely studied in the literature due to the relative ease of their precise manufacturing and developing numerical solutions to estimate their performance \cite{hopkins1999,yang2008,lips2009,lips2010,chen2014,kim2016,alijani2018a,alijani2018b}. Moreover, micro-electro-mechanical systems (MEMS) based micro-fabrication techniques enabled the fabrication of micro scale grooves on semiconductors, which opened an opportunity for direct integration of heat pipes onto electronic/optoelectronic chips \cite{peterson1993,harris2010,kundu2015}. Furthermore, extreme miniaturization in semiconductor devices may require thermal management solutions in the nanoscale. While the potential benefits of nano- and atomic-scale fluid flow (\textit{e.g.} large slip lengths, ballistic gas transport) \cite{majumder2005,holt2006,celebi2017,radha2018} and nanoscale ultra fast evaporation \cite{radha2016,li2017} promote the possible applications of nanoscale phase-change thermal management devices, the main hurdle seems to be the fabrication and integration of nanoscale capillaries. For example, carbon nanotubes (CNTs) provide exceptional structural, electrical, and thermal properties at nanoscale, but it is still difficult to integrate these superior properties in macroscopic devices \cite{radha2016}. On the other hand, recent fabrication techniques enabled production of nano- and \aa ngstr\"{o}m-scale  capillaries and cavities using top-down approach, which made the integration of nanoscale details onto larger scale devices possible \cite{radha2016,li2017,radha2018}.

At the nanoscale, liquids exhibit density fluctuations due to the wall-force-field effect. Apparent viscosity \cite{vo2015} and density \cite{ghorbanian2016} deviate from their bulk fluid properties. Vapor flow in nano-confinements require kinetic theory based modeling either in the transition or free molecular flow regimes. Influence of interfacial thermal resistance becomes important \cite{wang2007}, liquid-solid contact angle exhibits considerable variations \cite{barisik2013}, and capillary pumping mechanism resulting from the meniscus deformation may extinguish due to the annihilation of liquid meniscus structures \citep{akkus2018} at the nanoscale. Therefore, performance prediction of a nano-grooved heat pipe is not straightforward. Combined effect of the afore-mentioned nanoscale factors can be only addressed using atomistic level simulations.

Motivated by the recent advances in nano-fabrication techniques, extreme miniaturization trends in semiconductor industry, chip-level thermal management needs and potential benefits of nanoscale mass and energy transport, we investigate transport in nanoscale heat pipes with nano-grooves. Our ultimate objective is to develop a computational setup for performance characterization of a nano-grooved heat pipe. Using this setup, the effects of filling ratio and heat loads on the working performance are evaluated. Moreover, assessment of size effects on the working performance is performed by comparing proportionally scaled heat pipes. To the best knowledge of authors, current study presents the first computational work on \textit{nano-grooved heat pipes}. Previous attempts were using surfaces of a post wall \cite{moulod2016} and sharp-angled corners \cite{wang2007} as liquid arteries between two reservoirs. Moreover, assessment of design and operating parameters on the thermal performance of nanoscale heat pipes are investigated for the first time in the literature.

\section{Performance prediction in continuum scale}
\label{sec:pumping}

\begin{figure}[h]
	\centering	
	\epsfig{file=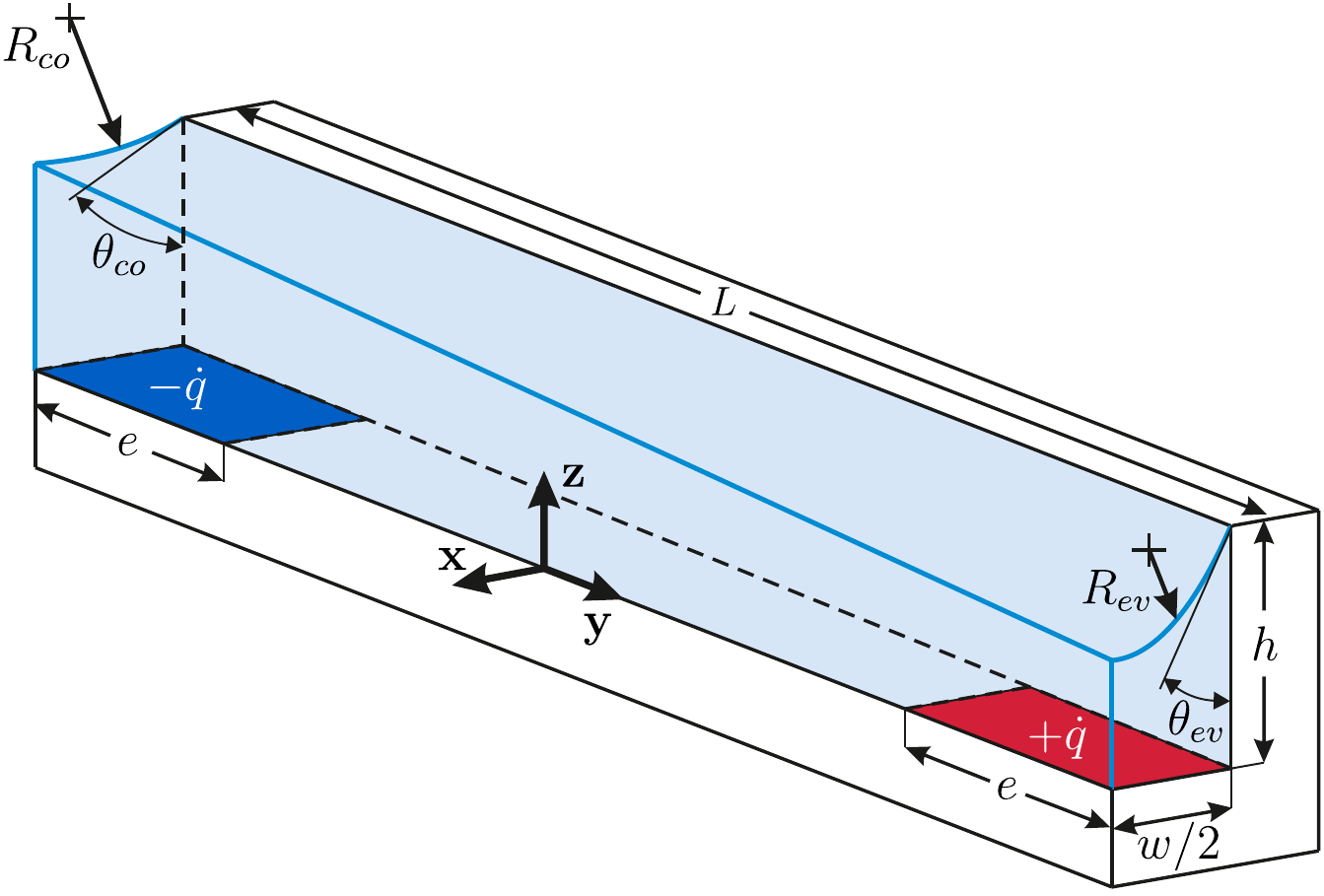,width=4in}
	\caption{\label{fig:half_groove}Single (half) heat pipe groove with geometrical and thermal parameters.}
\end{figure}

The current study evaluates the performance of proportionally scaled heat pipes in order to assess the size effect on the working characteristics. To perform such an analysis, geometric similarity between different sized heat pipes should be secured. Therefore, before starting computational experiments on different sized systems, we estimate the continuum scale performance of a grooved heat pipe by conducting a flow and evaporative mass analysis on a single half groove shown in Fig.~\ref{fig:half_groove}.  

The mass flow rate of a pressure induced laminar liquid flow in an open channel is proportional to the height, $h$, width, $w$, and the pressure gradient along the channel:

\begin{equation} \label{eq:mdot}
\dot m \sim\ h^2w^2\, \frac{dp}{dy} \ .  
\end{equation}

\noindent If the flow is driven by a capillary pressure gradient resulting from the asymmetric menisci at both ends of the groove, pressure gradient can be approximated using Young-Laplace equation:

\begin{equation}\label{eq:dpdx}
\frac{dp}{dy} \simeq \frac{\sigma}{L} \Bigl(  \frac{1}{R_{co}} -\frac{1}{R_{ev}}  \Bigr) \ ,  
\end{equation}

\noindent where $\sigma$ and $L$  are the surface tension coefficient and the effective length of the groove, respectively. Radius of curvature, $R$, can be written as the functions of apparent contact angle, $\theta$, and groove width, $w$, 

\begin{equation}\label{eq:rteta}
R= \frac{w/2}{\cos\theta } \ .
\end{equation}

\noindent Combination of Eqn.~\eqref{eq:mdot}, Eqn.~\eqref{eq:dpdx} and Eqn.~\eqref{eq:rteta} yields the mass flow rate, $\dot m$, as the functions of apparent contact angles formed at the condenser and evaporator regions:

\begin{equation}\label{eq:mdot_teta} 
\dot m \sim\ \frac{h^2 w}{L}  \Bigl[  \cos{{\theta}_{ev}} - \cos{{\theta}_{co}} \Bigr] \ .
\end{equation}

\noindent Heat pipes use the phase-change mechanism to remove heat from the source. Liquid turns into vapor by absorbing heat from the source at the evaporator section. Then vapor travels along the heat pipe and condenses back into liquid phase by releasing its energy in the form of latent heat at the condenser section. However, heat conduction through the base material, liquid, and vapor also takes place simultaneously. The efficacy of a heat pipe is preserved as long as the phase-change is the dominant heat transfer mode. Therefore, the ratio of energy transferred by phase-change, $\dot q_{p.c.}$, to the energy input at the evaporator, $\dot q$, reflects the efficiency of a heat pipe:

\begin{equation} \label{eq:eff}
\eta \equiv \frac{\dot q_{p.c.}}{\dot q} \ .
\end{equation}

\noindent The evaporating mass flow rate is equal to the liquid mass flow rate along the groove during the steady operation of the heat pipe. Therefore, the amount of energy removed from the liquid/vapor interface by phase-change mechanism can be written in terms of the liquid flow rate and heat of vaporization, $h_{\mathit{fg}}$, as follows:  

\begin{equation} \label{eq:q_pc}
\dot q_{p.c.} = \dot m h_{\mathit{fg}} \ .
\end{equation}

\noindent When Eqn.~\eqref{eq:mdot_teta} and Eqn.~\eqref{eq:q_pc} are inserted to Eqn.~\eqref{eq:eff}, efficiency of the heat pipe can be demonstrated as the functions of both heat inputs and system geometry. After some algebraic manipulations, the efficiency can be shown as follows: 

\begin{equation} \label{eq:nondim_long}
\eta \sim\  \Bigl( \frac{\dot q}{ew} \Bigr) ^{-1} \Bigl(\frac{e}{h} \Bigr)^{-1} \Bigl(\frac{L}{h}\Bigr)^{-1} \Bigl[  \cos{{\theta}_{ev}} - \cos{{\theta}_{co}}  \Bigr]
\end{equation}

\noindent where $e$ is the length of the region, where heat is added. The first term at the right hand side is simply the heat flux applied to the liquid at the evaporator, $\dot q''$. The second and third terms are scaled heat addition length, $e^*$, and aspect ratio of the groove, $L^*$, respectively. Therefore, Eqn.~\eqref{eq:nondim_long} can be written in terms of these parameters as follows:

\begin{equation} \label{eq:nondim_short}
\eta \sim\ (\dot q'')^{-1} \ \Bigl[ \frac{\cos{{\theta}_{ev}} - \cos{{\theta}_{co}} }{e^* L^*} \Bigr] \ .
\end{equation}

\noindent Eqn.~\eqref{eq:nondim_short} implies that if the heat flux applied and geometric similarity of the system are preserved, the heat pipe should exhibit similar performance. Therefore, while we are proportionally scaling down the system, we also keep the heat flux same by decreasing the heat input to the system during simulations. In other words, while we are decreasing the groove dimensions and heat input ($L$, $h$, $e$ and $\dot q$), we preserve the geometric and thermal similarity terms ($L^*$, $e^*$ and $\dot q''$). 

\section{Simulations}
\label{sec:sims}

\begin{figure}[h]
	\centering	
	\epsfig{file=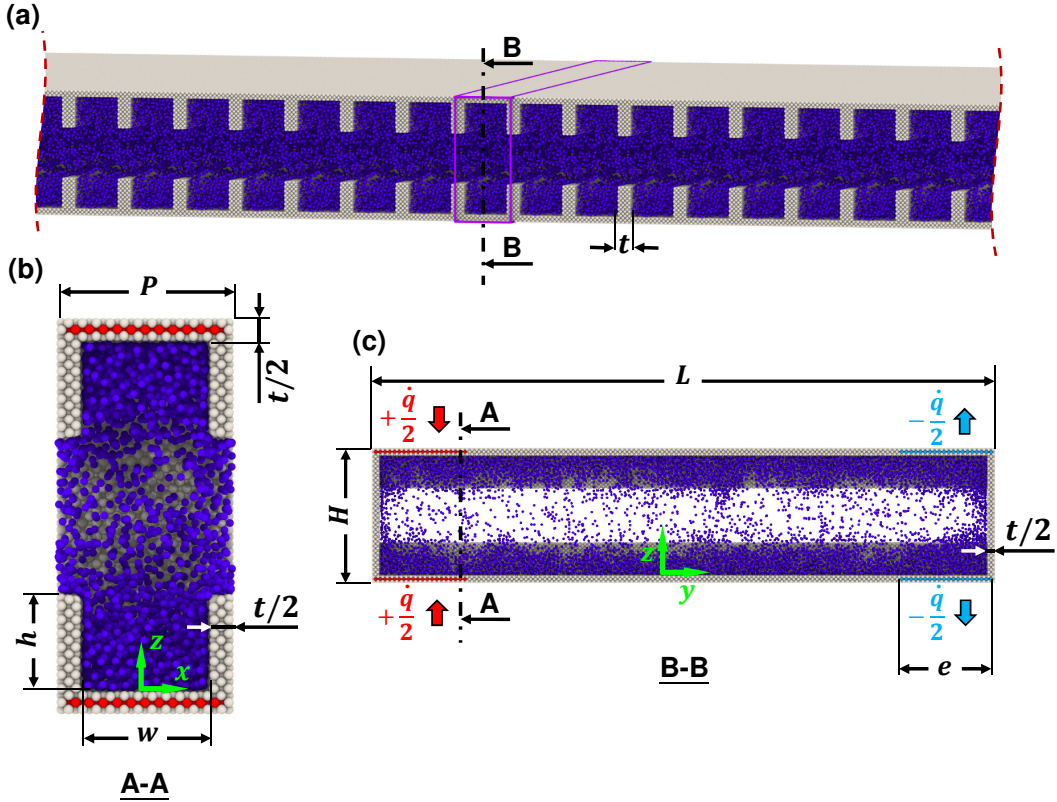,width=5.5in}
	\caption{\label{fig:domain}Schematic of the nano-grooved flat plate heat pipe composed of Platinum walls (gray spheres) and filled with Argon (blue spheres) as the working fluid. (a) A portion of the lateral cross-section of nano-grooved heat pipe. Purple lines indicate the computational domain. Heat pipe geometry is formed by the periodic images of the computational domain in $x$-direction. (b) Lateral cross-section of the computational domain. (c) Longitudinal cross-section of the computational domain. Red and turquoise spheres show the Platinum atoms subjected to energy injection and extraction at the evaporator and condenser sections, respectively.}
\end{figure}

Two proportionally scaled nano-grooved flat plate heat pipes are investigated. Heat pipes are modeled using molecular dynamics (MD) simulations. Schematic of the heat pipes is shown in Fig.~\ref{fig:domain}. The values of the dimensions in the schematic are specified in \mbox{Table~\ref{table:hp_dim}}. In general, a grooved flat plate heat pipe uses multiple axial grooves as liquid arteries to deliver the condensate to the evaporator. To create a nano-grooved heat pipe in MD environment, firstly, we construct a cell structure (the simulation domain) composed of two symmetric grooves with closed ends. Lateral and longitudinal cross sections of the cell structure are shown in Fig.~\ref{fig:domain}b and Fig.~\ref{fig:domain}c, respectively. Then, periodic boundary condition is applied in $x$-direction, which replicates the simulation domain throughout $x$-direction to form an infinite array of grooves, \textit{i.e.} the nano-grooved heat pipe, as shown in Fig.~\ref{fig:domain}a. Therefore, the cell structure is used as the simulation domain and computational experiments are carried out on this domain to evaluate the working characteristics and performance of the heat pipe.

The number of solid (Pt) atoms are 45,456 and 12,620 for \mbox{Heat Pipe-1} \mbox{(HP-1)} and \mbox{Heat Pipe-2} \mbox{(HP-2)}, respectively. To assess the effect of filling ratio, different amounts of fluid (Ar) are added to the heat pipes. While \mbox{HP-1} is filled with Ar within the range of 11,000--35,000 atoms, \mbox{HP-2} is tested with 1,600--4,800 Ar atoms. In each system, walls of the simulation domain are composed of 3 solid layers and (1,0,0) crystal planes face the fluid. The outermost layers of the walls are always fixed at their lattice positions. Time step is $5 \unit{fs}$ and each collected data is averaged for $1 \unit{ns}$. The interactions between \mbox{Ar-Ar} and \mbox{Ar-Pt} atoms are modeled using Lennard-Jones (L-J) 6-12 potential:

\begin{equation} \label{eq:lj}
\phi(r_{ij}) = 4 \varepsilon \Biggl[  \Bigl( \frac{\sigma}{r_{ij}} \Bigr)^{12} - \Bigl( \frac{\sigma}{r_{ij}} \Bigr)^{6}  \Biggr] \ .
\end{equation}

\noindent Molecular diameters and depth of the potential wells for the interactions are: $\sigma_{Ar}=0.34 \unit{nm}$, $\sigma_{Ar-Pt}=0.3085 \unit{nm}$ and $\varepsilon_{Ar}=0.01042 \unit{eV}$, $\varepsilon_{Ar-Pt}=0.00558 \unit{eV}$ \citep{maruyama1999}. L-J potential is truncated with a cut-off distance of $2.6\sigma_{Ar}$. Pt-Pt atomic interactions are modeled using embedded atom model \citep{foiles1986}. 

Simulations are started from the Maxwell-Boltzmann velocity distribution for all atoms at $110 \unit{K}$. First, Nos\'e-Hoover thermostat (NVT ensemble) is applied to all atoms (except the outermost atoms covering the surface of the walls) for $20 \unit{ns}$ to stabilize the system temperature at $110 \unit{K}$. Then, microcanonical (NVE) ensemble is applied to Ar atoms for $20 \unit{ns}$ to equilibrate the system, while wall atoms are still subjected to the thermostat. At the end of the equilibration period, stable liquid/vapor Ar mixture is obtained at $110 \unit{K}$. While saturated Argon condenses within grooves due to the interaction of fluid atoms with solid wall atoms, vapor phase of Argon occupies rest of the simulation domain.

\begin{table*} []
	\centering
	\caption{Dimensions of the heat pipes}
	\label{table:hp_dim}
	\begin{tabular}{lccccccc}\\
		\hline\\[-3pt]
		&\textbf{w [nm]}& \textbf{h [nm]}&\textbf{e [nm]}& \textbf{t [nm]} &\textbf{P [nm]}&\textbf{H [nm]}&\textbf{L [nm]}\\[5pt]
		\hline\\[-5pt]
		Heat Pipe-1 &3.920 &3.136 &7.840 &0.784 &4.704 &10.976 &51.744   \\
		Heat Pipe-2 &1.960 &1.568 &3.920 &0.784 &2.744 &5.880 &26.264   \\[3pt]
		\hline                            
	\end{tabular}
\end{table*}

Operation of the heat pipe is initiated with application of equal energy injection and extraction to the solid atoms located at the evaporator and condenser sections, respectively. During heating/cooling, Ar atoms are subjected to NVE ensemble. Heat transfer to/from the heat pipe is performed by energy injection/extraction from solid atoms instead of thermostat application. This method eliminates the non-physical temperature jump caused by thermostats \cite{barisik2012}. Moreover, zero net heat transfer to the heat pipe is secured by this approach. Before the collection of the data, heat pipe is allowed to operate for $60 \unit{ns}$, which is substantially longer than the diffusion time scales of momentum, $L^2/\nu$, and heat, $L^2/\alpha$, where $\nu$ and $\alpha$ are the kinematic viscosity and thermal diffusivity, respectively. Simulations are ceased at the end of $500 \unit{ns}$ heat operation. Statistically stable liquid/vapor interface profile ensures steady state fluid circulation within the heat pipe. All simulations are carried out using Large-scale Atomic/Molecular Massively Parallel Simulator (LAMMPS) \citep{plimpton1995}.

\subsection{Selection of working fluid}

Selection of working fluid is critical for proper functioning of a heat pipe. In order to create a capillary pumping mechanism, working fluid should be compatible with the wall material, \textit{i.e.} liquid phase must wet the solid surface \cite{faghri1995,reay2013}. Wettability of liquid Ar on Pt surface was confirmed in previous studies \cite{matsumoto1997,cao2006} for LJ potential parameters utilized in the current work. Another reason of the selection of Ar is its high vapor pressure. The number of vapor atoms in the gas phase should be large enough for accurate statistical averaging. Due to the computational cost of MD simulations, only a limited number of atoms/molecules can be simulated within restricted time spans. Therefore, selection of a fluid with high vapor pressure is inevitable. For instance, water, which is generally utilized in thermal engineering applications, is not a good choice for MD simulations due to its relatively low volatility. Saturated water mixture in equilibrium at room temperature has the ratio of $\sim1/50,000$ vapor to total water molecules, which is unfavorable for MD simulations conducted in nanoscale volumes. Saturated Ar, on the other hand, has the ratio of $\sim 1/40$ vapor to total Ar atoms at $110 \unit{K}$ temperature.

\section{Results and Discussions}
\label{sec:results}

\begin{figure}[t!]
	\centering	
	\epsfig{file=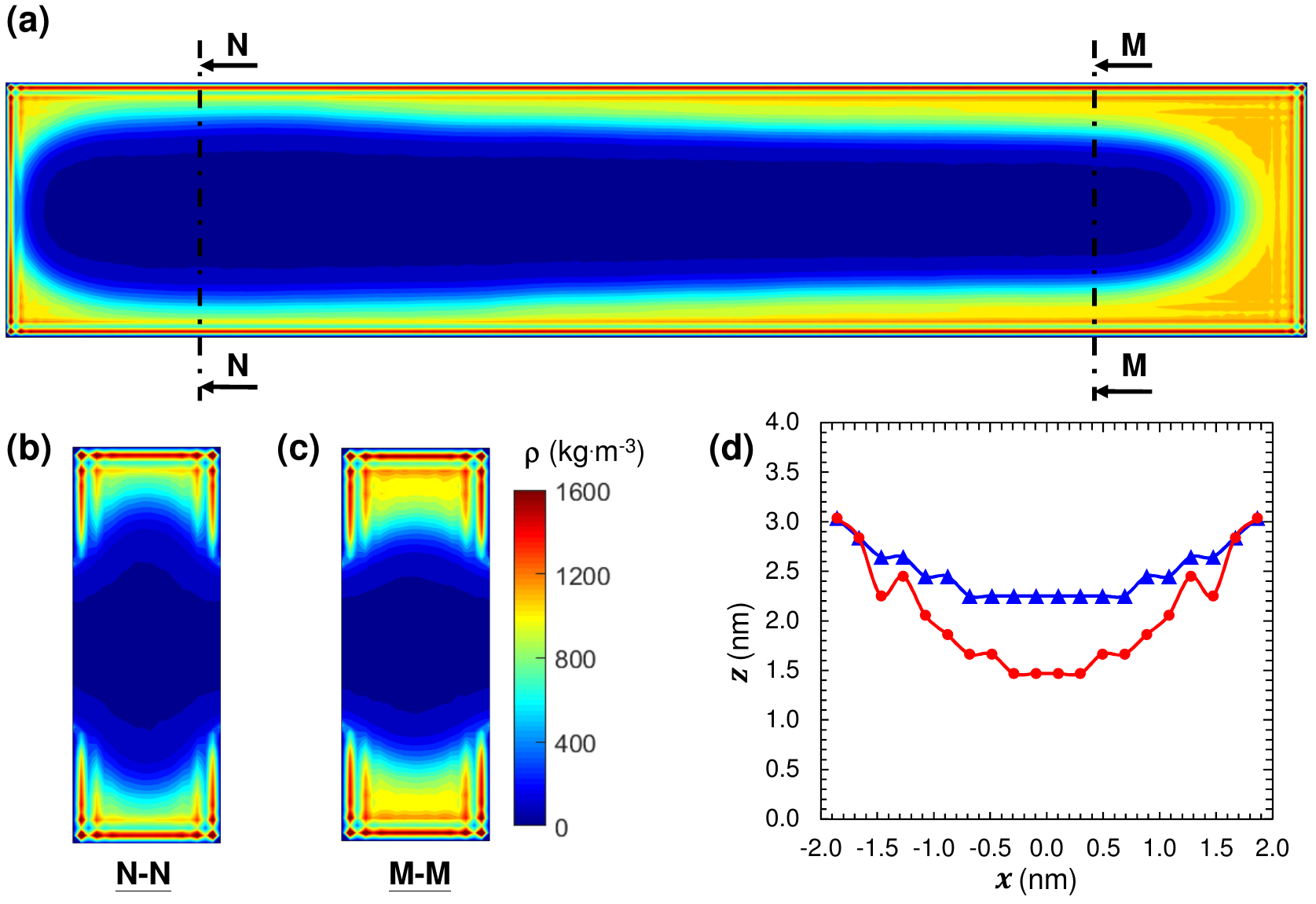,width=5.5in}
	\caption{Density distribution and interface profiles in \mbox{HP-1} under 1.0\unit{nW}heat load with the filling ratio of 0.43. (a) Density distribution along the groove axis on the central \textit{y-z} plane. Density distributions on \textit{x-z} plane near (b) evaporator and (c) condenser. Planes are $L/6$ distance away from heat pipe ends. (d) Liquid/vapor interface profiles near evaporator (circular marks) and condenser (triangular marks). Data points are connected with smooth lines to guide the eye. Starting from the outer gas phase region, density of each bin is checked in $z$-direction and the first bins, where the bin density exceeds the cut-off density ($375\unit{kg/m^3}$), are marked as the liquid/vapor interface. Cutoff density is selected to be slightly higher than the density of the bin at groove-fin top corner.}
	\label{fig:dens_v2_cond}
\end{figure}

Initial simulations are performed on HP-1 for different filling ratios. When all the inner space of the heat pipe is filled with the liquid phase, \textit{i.e.} fully flooded operation, the filling ratio is taken as unity. For the other cases, saturated liquid and vapor phases coexist, and the ratio of total mass of the two phases to the liquid mass during fully flooded operation defines the filling ratio of the heat pipe. Time-averaged density distribution in the heat pipe during a steady operation with filling ratio of 0.43 under 1.0\unit{nW}heat load is given in Fig.~\ref{fig:dens_v2_cond}a-c at different cross-sections. Variation of the liquid/vapor interface profile along the groove axis is demonstrated by the density distribution on the central \textit{y-z} plane (Fig.~\ref{fig:dens_v2_cond}a). Relative equilibrium amount of liquid is higher at the condenser due to mass accumulation. Near the side wall of the condenser, the liquid phase occupies the space between the grooves and makes a liquid bridge. This bridge has a circular interface with the vapor phase due to surface tension. At the evaporator region, the amount of liquid is not enough to form a continuous bridge, instead, two separate menisci form between the side wall and both grooves. The menisci are connected by an adsorbed liquid layer on the side wall surface. Liquid height at the center of the groove decreases gradually towards the evaporator. Distributions of the two phases between groove walls near the condenser and evaporator regions are shown in \mbox{Fig.~\ref{fig:dens_v2_cond}b~and~Fig.\ref{fig:dens_v2_cond}c}, respectively. Argon condenses within the grooves and vapor phase occupies the rest. Molecular layering of liquid is apparent near the side and outer walls with two distinct layers. The interface profile between liquid and vapor is determined based on a cutoff density and plotted for the both cases in Fig.~\ref{fig:dens_v2_cond}d. Both interfaces are attached to the edges of the sidewalls eliminating the possibility of a dryout or corner flow \cite{nilson2006}. Near the side walls, strong molecular layering of liquid affects the interface profile especially near the evaporator. Away from the walls, cohesion forces (surface tension) bend the interface and lead to the formation of meniscus structures. Meniscus is less apparent at the interface near condenser due to the thicker liquid film. Although curvature of each meniscus cannot be determined by a constant value due to the strong liquid-solid interaction, effective radius of curvatures of two menisci is appreciably different suggesting the presence of a capillary pumping mechanism between evaporator and condenser. Moreover, there exists a density difference between the liquid phases (especially visible in near-wall layers \cite{yin2017}), which indicates the presence of molecular diffusion together with the capillary pumping to convey the liquid from condenser to evaporator \cite{akkus2018}.                   

\begin{figure}[h]
	\centering	
	\epsfig{file=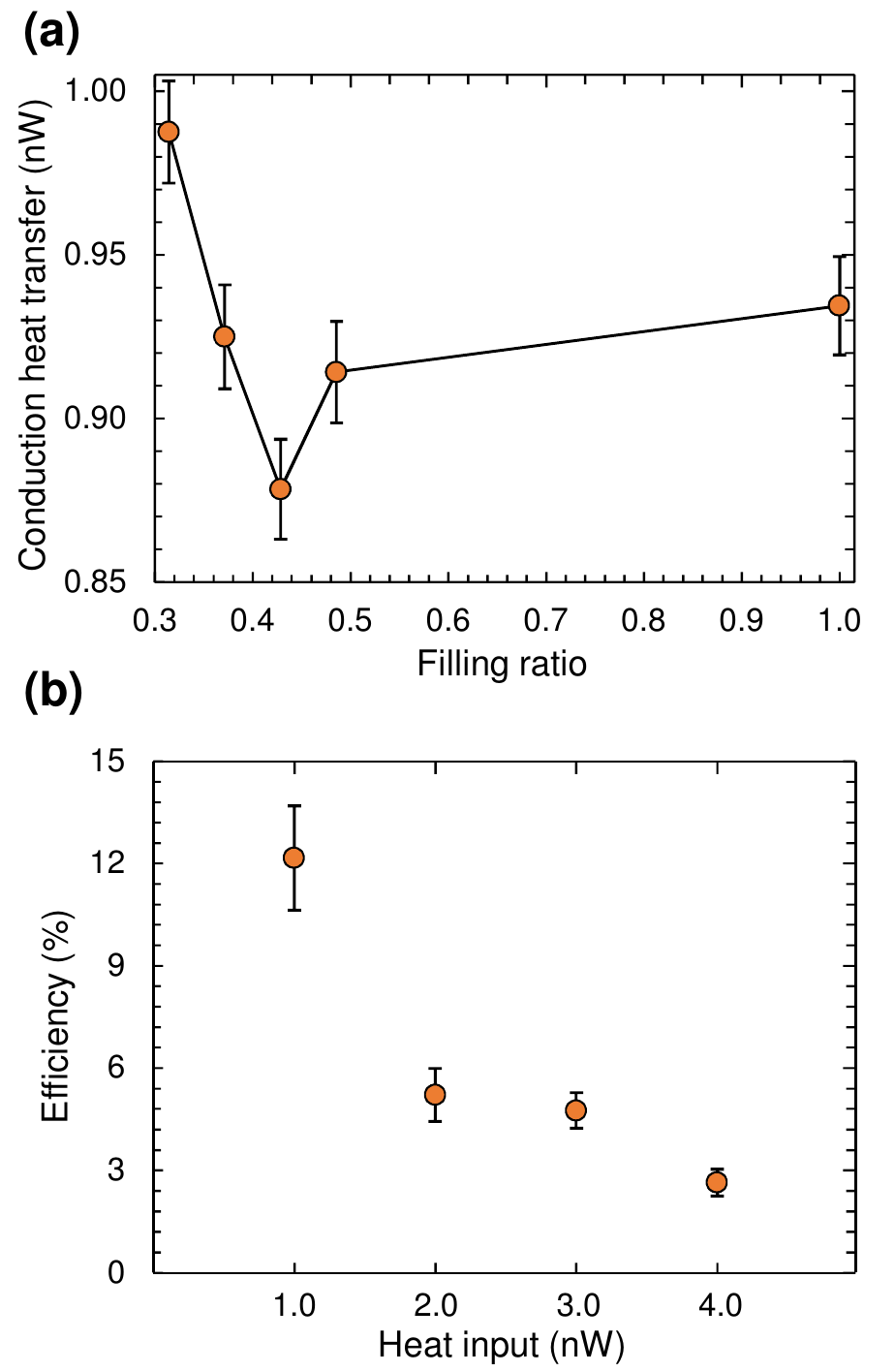,width=3.1in}
	\caption{Performance evaluation of \mbox{HP-1}. (a) Conduction heat transfer rate through walls vs. filling ratio under 1.0\unit{nW}heat load. (b) Heat pipe efficiency vs. heat input (load) for the filling ratio of 0.43.}
	\label{fig:conductive}
\end{figure}

\begin{figure}[t!]
	\centering	
	\epsfig{file=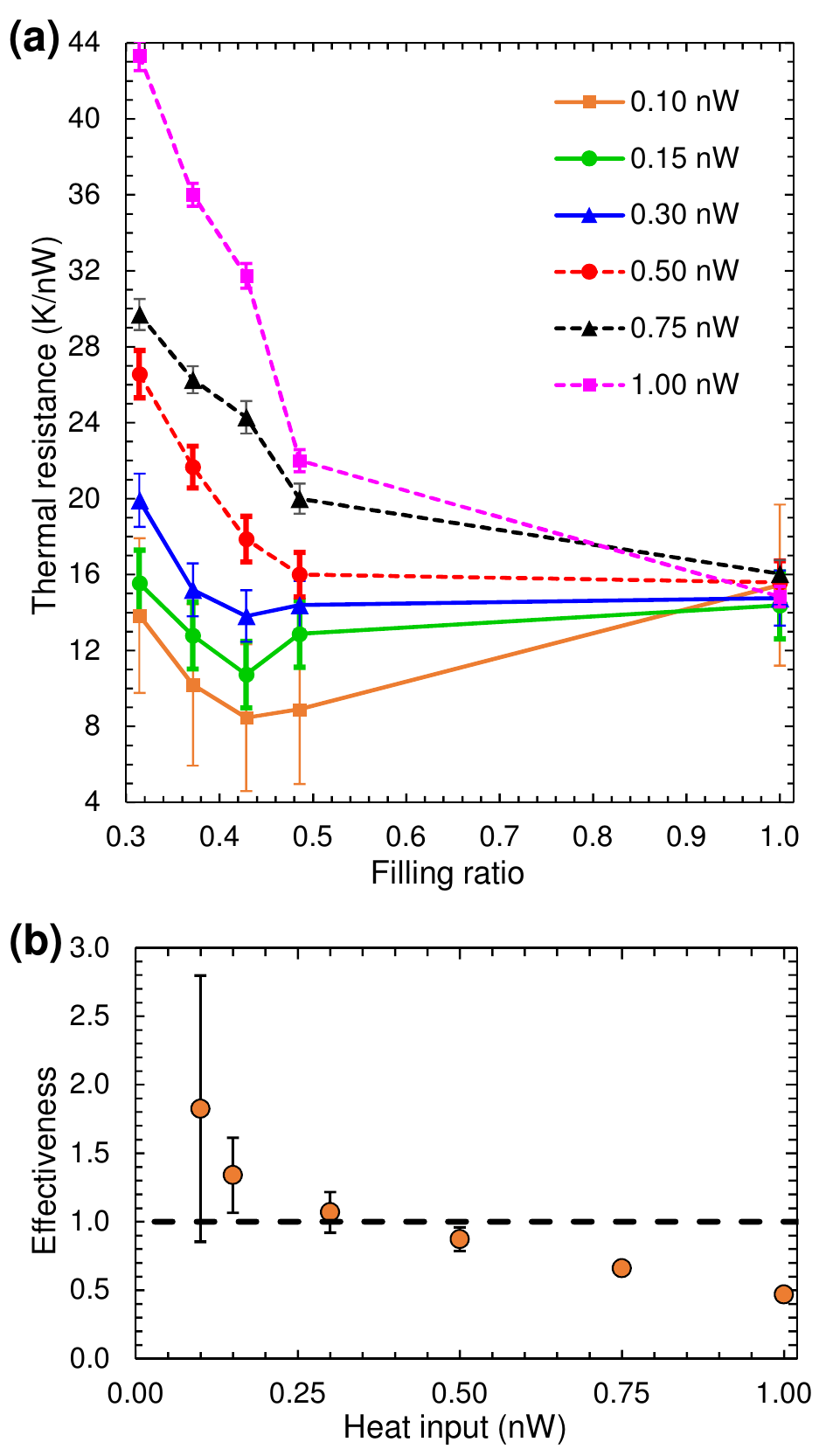,width=3in}
	\caption{Performance evaluation of \mbox{HP-1} with adiabatic walls between evaporator and condenser. (a) Thermal resistance vs. filling ratio for various heat inputs. (b) Effectiveness of the heat pipe operation vs. heat input.}
	\label{fig:v2_adb}
\end{figure}

Heat pipes use the advantage of phase-change to transfer thermal energy, which enables them to operate with small temperature differentials. Therefore, the ratio of energy transferred by the phase-change mechanism should be as high as possible. In other words, axial conduction should be minimized to realize an efficient operation with a small temperature gradient. Heat is conducted through both fluid and solid media within the heat pipe, but the latter dominates the former one due to the difference of thermal conductivity. The conduction heat transfer rate through solid walls between evaporator and condenser is calculated during simulations in order to make a performance assessment and its variation with the filling ratio is shown in \mbox{Fig.~\ref{fig:conductive}a} under a heat load of 1.0\unit{nW}. Similar its macro-scale experimental counterparts \cite{lips2009,lips2010,chen2014,alijani2018a,alijani2018b}, our computational results reveal that the nano-grooved heat pipe exhibits an optimum performance at a certain filling ratio. While the lesser amount of fluid causes an increase in conduction heat transfer due to the deficit of phase-changing fluid, which eventually leads to the occurrence of dryout (dry regions in evaporator), higher amount of fluid leads to the decrease in the rates of phase-change because of the higher thermal resistance of thicker liquid films at the evaporator and condenser \cite{holm1979,moosman1980,do2008,akkus2016,akkus2017}. The effect of heat load on the thermal performance is investigated and the heat pipe efficiencies ($\eta$), calculated based on the ratio of heat transfered by the phase-change mechanism to the total heat load (Eqn.~\eqref{eq:eff}), are presented in \mbox{Fig.~\ref{fig:conductive}b}. Performance of the heat pipes (and the the associated uncertainties) increase with smaller heat inputs in accordance with \cite{lips2009,alijani2018a}. However, efficiency of HP-1 is  appreciably low for all heat inputs, indicating a conduction heat transfer dominant operation. The reason of this behavior lies in the fact that aspect ratio of the grooves ($L/h$=16.5) is extremely small in the simulations with respect to that in conventional heat pipes ($\sim 10^2$--$10^4$). Thus, thermal resistance of conduction heat transfer is relatively low in HP-1. 

Simulation of a heat pipe with the groove aspect ratio similar to ones in conventional heat pipes is not possible due to the limits of the current computational power. However, conduction heat transfer through the walls can be prevented by manipulating the solid atoms. Therefore, instead of constructing a long heat pipe, we halted thermal vibrations of Pt atoms between the evaporator and condenser sections to mimic a sufficiently large conduction resistance such that all heat is transferred by phase-change mechanism and the conduction through fluid media. HP-1 with adiabatic mid section is simulated for different filling ratios under various heat inputs. We utilize the thermal resistance between evaporator and condenser sections ($R_{th}$) for the assessment of thermal performance \cite{hopkins1999,yang2008,lips2009,lips2010,chen2014,kim2016}. Temperatures of the evaporator ($T_{ev}$) and condenser ($T_{co}$) sections are calculated based on the average temperature of Pt atoms subjected to energy injection and extraction, respectively. Temperature difference ($\Delta T=T_{ev}-T_{co}$) is divided by heat input to calculate the thermal resistance, \textit{i.e.}, $R_{th}=\Delta T/ \dot q$. Thermal resistance variations of the heat pipe with filling ratios under different heat inputs are shown in \mbox{Fig.~\ref{fig:v2_adb}a}. All the curves intersect at the fully flooded case, where the heat is transfered only \textit{via} conduction through the liquid phase, and the effective conductivity of the heat pipe is almost same regardless of the heat load applied. For the dry case, thermal resistance would be infinite due to the absence of any mechanism to convey the heat. Therefore, every curve extends between the same limits in \mbox{Fig.~\ref{fig:v2_adb}a}. However, these curves fall into two different groups in terms of their variation trends. For smaller heat inputs (0.10\unit{nW,}0.15\unit{nW,}0.30\unit{nW),}each simulation reveals an optimum operating point similar to the operation of HP-1 with conductive mid section. The optimum filling ratio was the same for HP-1, 0.43, regardless of the heat input \cite{chen2014,alijani2018a}. For higher heat inputs (0.50\unit{nW,}0.75\unit{nW,}1.00\unit{nW),}on the other hand, minimum thermal resistance is achieved at the filling ratio of 1.0 without exhibiting any dip between dry and fully flooded operations. This behavior shows that majority of the heat is transferred by axial conduction through the fluid media for higher heat loads. Therefore, heat pipe does not function properly, and the benefit gained by phase-change becomes restricted at elevated heat inputs, \textit{i.e.}, heat pipe is overloaded. In order to draw a clearer picture on the efficacy of heat pipe operations, we utilize an effectiveness parameter ($\varepsilon$) based on the ratio of thermal resistance during fully flooded operation ($R_{th}^{\textit{\textsuperscript{f.f.}}}$) to that of heat pipe during optimum phase-change operation ($R_{th}^{opt.}$), \textit{i.e.}, $\varepsilon=R_{th}^{\textit{\textsuperscript{f.f.}}} / R_{th}^{opt.}$. The effectiveness is a function of the heat input and filling ratio for a given design \cite{alijani2018a}. The effectiveness values for \mbox{HP-1} with adiabatic mid section are shown in \mbox{Fig.~\ref{fig:v2_adb}b}, and an effectiveness value smaller than unity implies that heat transfer in the device is not dominated by phase-change. On the other hand, effectiveness of the heat pipe operation increases with diminishing heat loads. However, operational effectiveness should not be interpreted as the sole factor determining the general efficacy of the heat pipe. Further reduction in heat input may improve the effectiveness value by decreasing the temperature difference between evaporator and condenser, but the amount of heat removal would be less in that case. Therefore, general efficacy of the heat pipe is actually determined by the compromise between the amount of waste heat desired to be removed and the allowable temperature difference between heat source and external coolant.   

\begin{figure}[t!]
	\centering	
	\epsfig{file=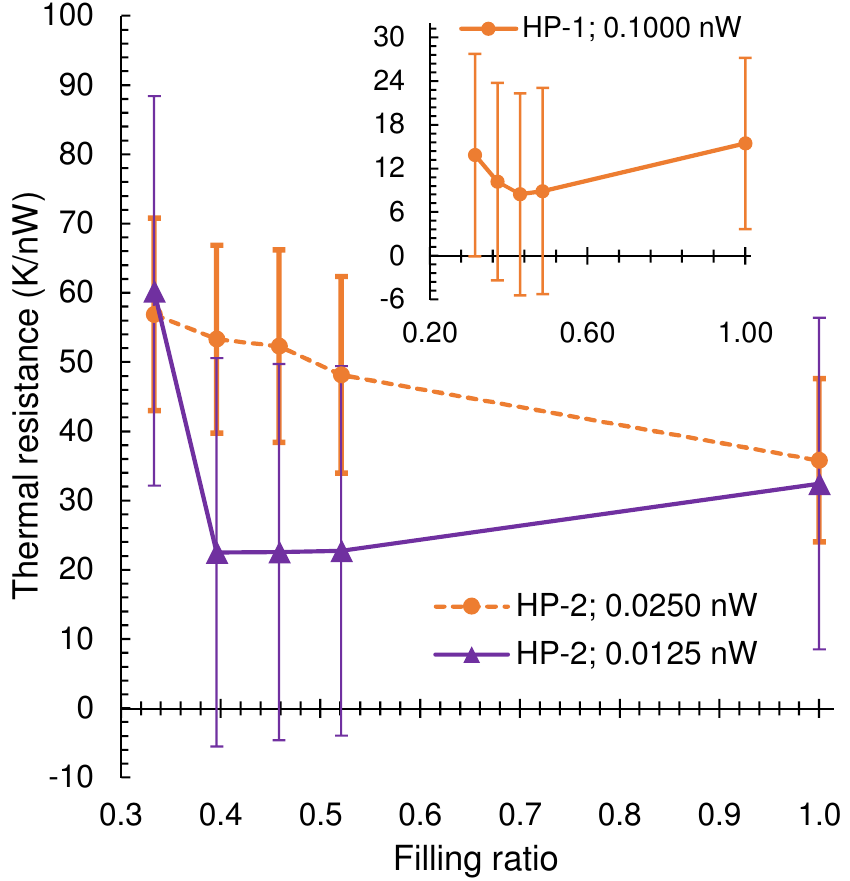,width=3.3in}
	\caption{Thermal resistance vs. filling ratio for \mbox{HP-2} for different heat inputs. Inset shows the thermal resistance variation \mbox{HP-1} for the heat input of 0.1000\unit{nW.} Operation of \mbox{HP-2} shown by orange dashed line has the same heat flux input with the operation of \mbox{HP-1} shown by the inset.}
	\label{fig:v2s_adb}
\end{figure}

\mbox{HP-1} is scaled down by a factor of 2 (\mbox{HP-2}) to assess the size effect on the working performance. The most effective operation of \mbox{HP-1} (with adiabatic mid section), which occurs under the heat load of 0.1000\unit{nW,}is compared with the operation of \mbox{HP-2} (with adiabatic mid section) for the same heat flux input, which corresponds to the heat load of 0.0250\unit{nW}for \mbox{HP-2}. Variations of thermal resistances with the filling ratio are given in  Fig.~\ref{fig:v2s_adb} (dashed line) and its inset for the operations of \mbox{HP-2} and \mbox{HP-1}, respectively. In the absence of nanoscale effects, continuum theory would yield a similar thermal performance for both heat pipes in accordance with Eqn.~\eqref{eq:nondim_short}. However, \mbox{HP-2} does not work effectively at the same heat flux as shown in Fig.~\ref{fig:v2s_adb}. We also halved the heat input of \mbox{HP-2} to observe its response to the reduction of heat load, and \mbox{HP-2} operates effectively under the reduced heat load (solid line in Fig.~\ref{fig:v2s_adb}). These results demonstrate that size reduction severely reduces the thermal performance of \mbox{HP-2}.

\begin{figure}[t!]
	\centering	
	\epsfig{file=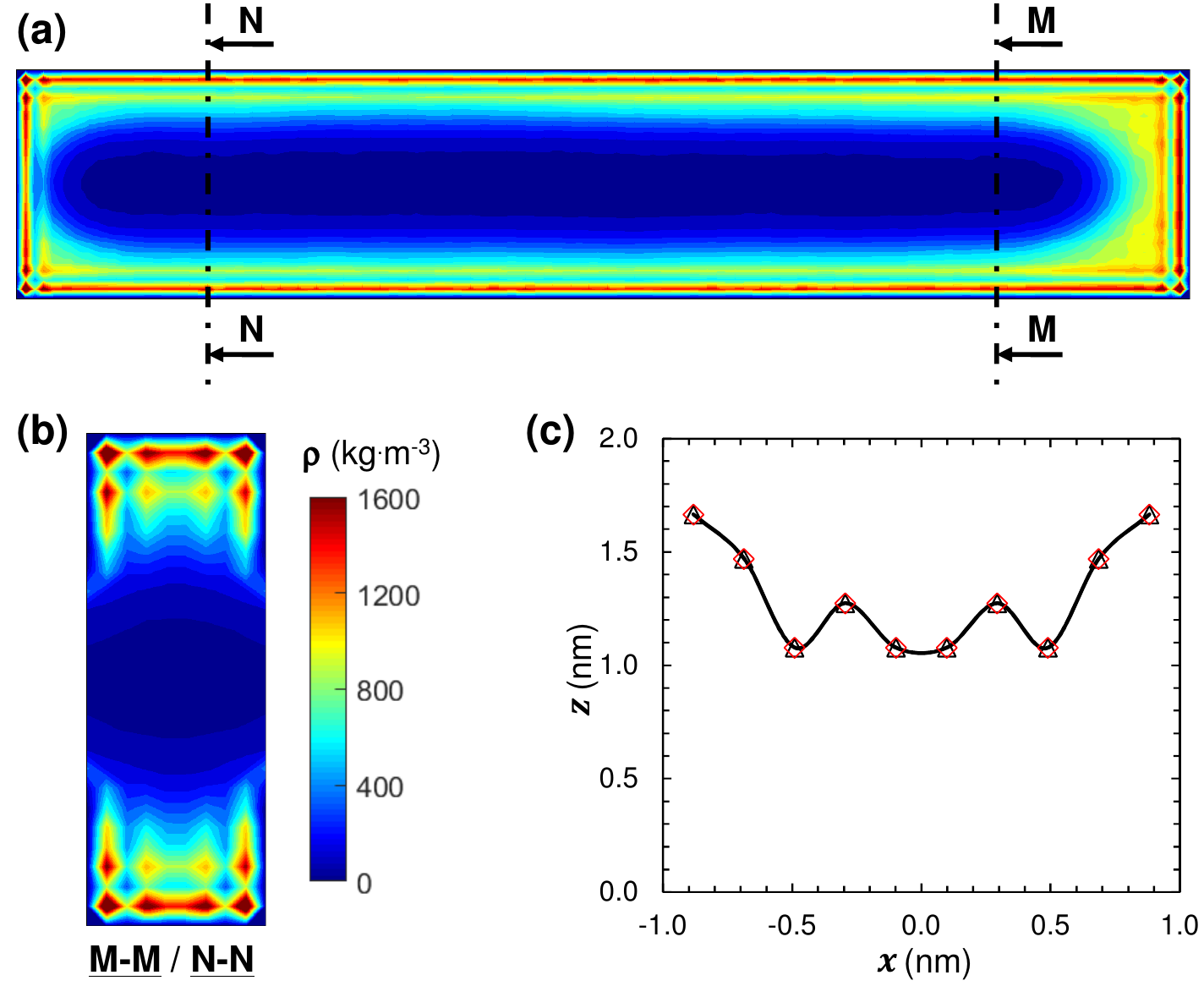,width=5in}
	\caption{Density distribution and interface profile in \mbox{HP-2} with adiabatic mid section under 0.0250\unit{nW}heat load with the filling ratio of 0.396. (a) Density distribution along the groove axis on the central $y$-$z$ plane. (b) Density distribution on $x$-$z$ plane near evaporator/condenser. Density distribution is almost same within the mid section between evaporator and condenser. Therefore, single plot is provided. (c) Liquid/vapor interface profiles near evaporator (diamond) and condenser (triangle). Interface profiles are exactly same. Data points are connected with smooth lines to guide the eye. Starting from the outer gas phase region, density of each bin is checked in $z$-direction and the first bins, where the bin density exceeds the cut-off density ($365\unit{kg/m^3}$), are marked as the liquid/vapor interface. Cutoff density is selected to be slightly higher than the density of the bin at groove-fin top corner.}
\label{fig:dens_v2s_adb}
\end{figure}

In order to determine the reason of performance loss due to scale reduction, time-averaged density distribution within \mbox{HP-2} is investigated. Variation of the liquid/vapor interface profile along the groove axis is demonstrated by the density distribution on the central $y$-$z$ plane (Fig.~\ref{fig:dens_v2s_adb}a). Unlike \mbox{HP-1}, where height of the liquid phase decreases gradually towards the evaporator, liquid height is constant between the evaporator and condenser in \mbox{HP-2}. Moreover, density distribution of the two phases between groove walls (\mbox{Fig.~\ref{fig:dens_v2s_adb}b}) is almost same along the mid region, \textit{i.e.} the region between evaporator and condenser sections. Furthermore, the interface profile between the liquid and vapor phases is exactly same at any $x$-$z$ plane lying within the mid region as shown in \mbox{Fig.~\ref{fig:dens_v2s_adb}c}, where the strong liquid-solid interaction severely affects the liquid film profile. The free surface between liquid and vapor is not able to form meniscus shaped interfaces with different curvatures along the groove axis. Therefore, capillary pumping mechanism between the evaporator and condenser vanishes, and molecular diffusion resulting from the different densities of the liquid at the evaporator and condenser sections provide the liquid transport \cite{akkus2018}. However, molecular diffusion is unable to maintain the thermal performance of the heat pipe in absence of capillary pumping assistance.

\section{Conclusions}

Utilizing molecular dynamics, we constructed a computational setup for the assessment of thermal performance of a nano-grooved Platinum heat pipe filled with Argon as the working fluid. The main conclusions of the current study can be summarized as follows:

\begin{itemize} 
\item Similar to their macro-scale counterparts, nano-grooved heat pipes exhibit an optimum thermal performance at a certain filling ratio.
\item When nano-grooved heat pipes are thermally overloaded, heat is mainly transferred by axial conduction, which makes them ineffective for thermal management.
\item As long as the effective operation of the nano-grooved heat pipe is realized, the optimum filling ratio is the same for different heat inputs.
\item When the size of nano-groove is $3.920\unit{nm}\times 3.136\unit{nm}$, both capillary pumping and molecular diffusion exist to transport the condensate from condenser to evaporator.
\item Scaling down the system (a nano-groove size of $1.960\unit{nm} \times 1.568\unit{nm}$) leads to thermal performance loss resulting from the vanishing of capillary pumping due to the strong liquid/solid interaction forces that dominate the free surface. This prevents the formation of meniscus shaped interface under the effect of cohesion forces.
\end{itemize} 

\noindent As the final remark, simulations of the current study demonstrate that a nano-grooved heat pipe (\textit{e.g.} \mbox{HP-1} with adiabatic mid section) can operate under tremendous heat flux values, $\dot{q}''=\dot{q}/(ew)$, as high as  $\sim980\unit{W/cm^2}$. However, aspect ratio of the heat pipe grooves was kept restricted during simulations due to the limitation of the computational power. On the other hand, recent fabrication techniques enabled production of nano- and \aa ngstr\"{o}m-scale capillaries and cavities with higher aspect ratios \cite{radha2016,li2017,radha2018}, which motivates researchers to elucidate the thermal performance of nanoscale heat pipes with higher aspect ratios. If the length of the heat pipe is increased, viscous losses associated with liquid and vapor flows will rise. This, in turn, will result in decreased heat removal capacity of the heat pipe. Our future research will focus on the investigation of heat removal capacity of the nanoscale systems with simpler geometries but higher aspect ratios.

\section*{Acknowledgments}
\addcontentsline{toc}{section}{Acknowledgements}
Y.A. acknowledges the financial support of ASELSAN Inc. under scholarship program for postgraduate studies. Computations were carried out using high performance computing facilities of Center for Scientific Computation at Southern Methodist University.


\section*{Author contributions}
\addcontentsline{toc}{section}{Author_contributions}
Y.A. and C.T.N. performed molecular dynamics simulations. Y.A. and A.T.C. wrote the manuscript. All authors contributed most of the ideas and discussed the results. A.B. reviewed and edited the manuscript.

\section*{Competing interests}
\addcontentsline{toc}{section}{Competing_interests}
The authors declare no competing financial interests.

\section*{Appendix A. Uncertainty analysis}
\addcontentsline{toc}{section}{Competing_interests}

During the study, heat pipe efficiency, thermal resistance, and effectiveness are utilized for the assessment of thermal performance. While efficiency is calculated based on the measured conduction heat transfer rate ($\dot{q}_{cond}$) through the mid section, thermal resistance and effectiveness require the measurements of temperature at the evaporator and condenser sections. Conduction rate and temperature are sampled at every $1\unit{ns}$ and all the data collected between two measurements is averaged, which yields a certain measurement uncertainty, $u$, for each time averaged data, $ \langle..\rangle $. 

\begin{equation} \label{eq:uncert_q}
\dot{q}_{cond} = \langle \dot{q}_{cond} \rangle \pm u_{\dot{q}_{cond}} \ ,
\tag{A.1a}  
\end{equation}
\begin{equation} \label{eq:uncert_T}
T = \langle T \rangle \pm u_{T} \ .
\tag{A.1b}  
\end{equation}

\noindent Uncertainties associated with conduction rate ($u_{\dot{q}_{cond}}$) and temperature ($u_{T}$) are estimated by calculating the standard error of measurements, which is evaluated by dividing the standard deviation of measurements to the number of samples. In the calculation of efficiency, uncertainty is only associated with $u_{\dot{q}_{cond}}$. However, during the calculation of thermal resistance and effectiveness, uncertainties of the temperature difference between the evaporator and condenser should be considered. Uncertainty of the temperature difference ($u_{\Delta T}$) is expressed in terms of the uncertainties of the evaporator ($u_{T_{ev}}$) and condenser ($u_{T_{co}}$) temperatures:

\begin{equation} \label{eq:uncert_rho}
\Delta T = \langle \Delta T \rangle \pm u_{\Delta T} \ ,
\tag{A.2a}  
\end{equation}
\begin{equation} \label{eq:uncert_r}
u_{\Delta T} = \sqrt{(u_{T_{ev}})^2+(u_{T_{co}})^2} \ .
\tag{A.2b}  
\end{equation}

\noindent While the uncertainty of thermal resistance is determined only by $u_{\Delta T}$, effectiveness has the uncertainty ($u_{\varepsilon}$) resulting from the uncertainties of the temperature differences evaluated at both optimum ($u_{\Delta T^{\textit{\textsuperscript{f.f.}}}}$) and fully flooded ($u_{\Delta T^{opt.}}$) conditions:

\begin{equation} \label{eq:uncert_effct}
\varepsilon = \langle \varepsilon \rangle \pm u_{\varepsilon} \ ,
\tag{A.3a}  
\end{equation}

\begin{equation} \label{eq:uncert_effctvnss}
u_{\varepsilon} = \langle \varepsilon \rangle  \sqrt{  
	\Biggl(   \frac{u_{\Delta T^{\textit{\textsuperscript{f.f.}}}}}{ \langle \Delta T^{\textit{\textsuperscript{f.f.}}} \rangle   }  \Biggr)^2 
	+	\Biggl(   \frac{u_{\Delta T^{opt.}}}{   \langle \Delta T^{\textit{opt.}} \rangle   } \Biggr)^2 
} \ .
\tag{A.3b}  
\end{equation}

\addcontentsline{toc}{section}{References}
\bibliographystyle{unsrt}
\bibliography{references}

\begin{thebibliography}{10}

\bibitem{moore1965}
G.E. Moore.
\newblock Cramming more components onto integrated circuits.
\newblock {\em Electronics}, 38(8):114, 1965.

\bibitem{calame2007}
J.P. Calame, R.E. Myers, S.C. Binari, F.N. Wood, and M.~Garven.
\newblock Experimental investigation of microchannel coolers for the high heat
  flux thermal management of {GaN-on-SiC} semiconductor devices.
\newblock {\em Int. J. Heat Mass Tran.}, 50(23-24):4767--4779, 2007.

\bibitem{bagnall2013}
K.R. Bagnall.
\newblock {\em Device-level thermal analysis of {GaN-based} electronics}.
\newblock PhD thesis, Massachusetts Institute of Technology, 2013.

\bibitem{qu2017}
J.~Qu, H.~Wu, P.~Cheng, Q.~Wang, and Q.~Sun.
\newblock Recent advances in {MEMS-based} micro heat pipes.
\newblock {\em Int. J. Heat Mass Tran.}, 110:294--313, 2017.

\bibitem{perpina2011}
X.~Perpi{\~n}{\`a}, X.~Jord{\`a}, M.~Vellvehi, and J.~Altet.
\newblock Hot spot analysis in integrated circuit substrates by laser mirage
  effect.
\newblock {\em Appl. Phys. Lett.}, 98(16):164104, 2011.

\bibitem{tavakkoli2016}
F.~Tavakkoli, S.~Ebrahimi, S.~Wang, and K.~Vafai.
\newblock Analysis of critical thermal issues in 3d integrated circuits.
\newblock {\em Int. J. Heat Mass Tran.}, 97:337--352, 2016.

\bibitem{prasher2005}
R.S. Prasher, J.Y. Chang, I.~Sauciuc, S.~Narasimhan, D.~Chau, G.~Chrysler,
  A.~Myers, S.~Prstic, and C.~Hu.
\newblock Nano and micro technology-based next-generation package-level cooling
  solutions.
\newblock {\em Intel Technology Journal}, 9(4), 2005.

\bibitem{grover1964}
G.M. Grover, T.P. Cotter, and G.F. Erickson.
\newblock Structures of very high thermal conductance.
\newblock {\em J. Appl. Phys.}, 35(6):1990--1991, 1964.

\bibitem{cotter1984}
T.P. Cotter.
\newblock Principles and prospects for micro heat pipes.
\newblock In {\em 5th International Heat Pipe Conference, Tsukuba, Japan},
  pages 328--334, 1984.

\bibitem{hopkins1999}
R.~Hopkins, A.~Faghri, and D.~Khrustalev.
\newblock Flat miniature heat pipes with micro capillary grooves.
\newblock {\em J. Heat Transf.}, 121(1):102--109, 1999.

\bibitem{yang2008}
X.F. Yang, Z.H. Liu, and J.~Zhao.
\newblock Heat transfer performance of a horizontal micro-grooved heat pipe
  using {CuO} nanofluid.
\newblock {\em J. Micromech. Microeng.}, 18(3):035038, 2008.

\bibitem{lips2009}
S.~Lips, F.~Lef{\`e}vre, and J.~Bonjour.
\newblock Nucleate boiling in a flat grooved heat pipe.
\newblock {\em Int. J. Therm. Sci.}, 48(7):1273--1278, 2009.

\bibitem{lips2010}
S.~Lips, F.~Lef{\`e}vre, and J.~Bonjour.
\newblock Combined effects of the filling ratio and the vapour space thickness
  on the performance of a flat plate heat pipe.
\newblock {\em Int. J. Heat Mass Tran.}, 53(4):694--702, 2010.

\bibitem{chen2014}
J.S. Chen and J.H. Chou.
\newblock Cooling performance of flat plate heat pipes with different liquid
  filling ratios.
\newblock {\em Int. J. Heat Mass Tran.}, 77:874--882, 2014.

\bibitem{kim2016}
H.J. Kim, S.H. Lee, S.B. Kim, and S.P. Jang.
\newblock The effect of nanoparticle shape on the thermal resistance of a
  flat-plate heat pipe using acetone-based {Al2O3} nanofluids.
\newblock {\em Int. J. Heat Mass Tran.}, 92:572--577, 2016.

\bibitem{alijani2018a}
H.~Alijani, B.~Cetin, Y.~Akkus, and Z.~Dursunkaya.
\newblock Effect of design and operating parameters on the thermal performance
  of flat grooved heat pipes.
\newblock {\em Appl. Therm. Eng.}, 132:174--187, 2018.

\bibitem{alijani2018b}
H.~Alijani, B.~Cetin, Y.~Akkus, and Z.~Dursunkaya.
\newblock Experimental thermal performance characterization of flat grooved
  heat pipes.
\newblock {\em Heat Transfer Eng.}, in press, 2018.

\bibitem{peterson1993}
G.P. Peterson, A.B. Duncan, and M.H. Weichold.
\newblock Experimental investigation of micro heat pipes fabricated in silicon
  wafers.
\newblock {\em J. Heat Transf.}, 115(3):751--756, 1993.

\bibitem{harris2010}
D.K. Harris, A.~Palkar, G.~Wonacott, R.~Dean, and F.~Simionescu.
\newblock An experimental investigation in the performance of water-filled
  silicon microheat pipe arrays.
\newblock {\em J. Electron. Packaging}, 132(2):021005, 2010.

\bibitem{kundu2015}
P.K. Kundu, S.~Mondal, S.~Chakraborty, and S.~DasGupta.
\newblock Experimental and theoretical evaluation of on-chip micro heat pipe.
\newblock {\em Nanosc. Microsc. Therm.}, 19(1):75--93, 2015.

\bibitem{majumder2005}
M.~Majumder, N.~Chopra, R.~Andrews, and B.J. Hinds.
\newblock Nanoscale hydrodynamics: enhanced flow in carbon nanotubes.
\newblock {\em Nature}, 438(7064):44, 2005.

\bibitem{holt2006}
J.K. Holt, H.G. Park, Y.~Wang, M.~Stadermann, A.B. Artyukhin, C.P.
  Grigoropoulos, A.~Noy, and O.~Bakajin.
\newblock Fast mass transport through sub-2-nanometer carbon nanotubes.
\newblock {\em Science}, 312(5776):1034--1037, 2006.

\bibitem{celebi2017}
A.T. Celebi, M.~Barisik, and A.~Beskok.
\newblock Electric field controlled transport of water in graphene
  nano-channels.
\newblock {\em J. Chem. Phys.}, 147(16):164311, 2017.

\bibitem{radha2018}
A.~Keerthi, A.K. Geim, A.~Janardanan, A.P. Rooney, A.~Esfandiar, S.~Hu, S.A.
  Dar, I.V. Grigorieva, S.J. Haigh, F.C. Wang, and B.~Radha.
\newblock Ballistic molecular transport through two-dimensional channels.
\newblock {\em Nature}, 558:420--424, 2018.

\bibitem{radha2016}
B.~Radha, A.~Esfandiar, F.C. Wang, A.P. Rooney, K.~Gopinadhan, A.~Keerthi,
  A.~Mishchenko, A.~Janardanan, P.~Blake, L.~Fumagalli, M.~Lozada-Hidalgo,
  S.~Garaj, S.J. Haigh, I.V. Grigorieva, H.A. Wu, and A.K. Geim.
\newblock Molecular transport through capillaries made with atomic-scale
  precision.
\newblock {\em Nature}, 538(7624):222, 2016.

\bibitem{li2017}
Y.~Li, M.A. Alibakhshi, Y.~Zhao, and C.~Duan.
\newblock Exploring ultimate water capillary evaporation in nanoscale conduits.
\newblock {\em Nano Lett.}, 17(8):4813--4819, 2017.

\bibitem{vo2015}
T.Q. Vo, M.~Barisik, and B.H. Kim.
\newblock Near-surface viscosity effects on capillary rise of water in
  nanotubes.
\newblock {\em Phys. Rev. E}, 92(5):053009, 2015.

\bibitem{ghorbanian2016}
J.~Ghorbanian, A.T. Celebi, and A.~Beskok.
\newblock A phenomenological continuum model for force-driven nano-channel
  liquid flows.
\newblock {\em J. Chem. Phys.}, 145(18):184109, 2016.

\bibitem{wang2007}
C.S. Wang, J.S. Chen, J.~Shiomi, and S.~Maruyama.
\newblock A study on the thermal resistance over solid--liquid--vapor
  interfaces in a finite-space by a molecular dynamics method.
\newblock {\em Int. J. Therm. Sci.}, 46(12):1203--1210, 2007.

\bibitem{barisik2013}
M.~Barisik and A.~Beskok.
\newblock Wetting characterisation of silicon (1, 0, 0) surface.
\newblock {\em Mol. Simulat.}, 39(9):700--709, 2013.

\bibitem{akkus2018}
Y.~Akkus and A.~Beskok.
\newblock Molecular diffusion replaces capillary pumping in phase-change driven
  nanopumps.
\newblock {\em arXiv preprint arXiv:1804.06056}, 2018.

\bibitem{moulod2016}
M.~Moulod and G.~Hwang.
\newblock Nano heat pipe: nonequilibrium molecular dynamics simulation.
\newblock In {\em ASME 2016 International Mechanical Engineering Congress and
  Exposition}, pages V008T10A021--V008T10A021. American Society of Mechanical
  Engineers, 2016.

\bibitem{maruyama1999}
S.~Maruyama and T.~Kimura.
\newblock A study on thermal resistance over a solid-liquid interface by the
  molecular dynamics method.
\newblock {\em Therm. Sci. Eng.}, 7(1):63--68, 1999.

\bibitem{foiles1986}
S.M. Foiles, M.I. Baskes, and M.S. Daw.
\newblock Embedded-atom-method functions for the fcc metals {C}u, {A}g, {A}u,
  {N}i, {P}d, {P}t, and their alloys.
\newblock {\em Phys. Rev. B}, 33(12):7983, 1986.

\bibitem{barisik2012}
M.~Barisik and A.~Beskok.
\newblock Boundary treatment effects on molecular dynamics simulations of
  interface thermal resistance.
\newblock {\em J. Comput. Phys.}, 231(23):7881--7892, 2012.

\bibitem{plimpton1995}
S.~Plimpton.
\newblock Fast parallel algorithms for short-range molecular dynamics.
\newblock {\em J. Comput. Phys.}, 117(1):1--19, 1995.

\bibitem{faghri1995}
A.~Faghri.
\newblock {\em Heat {P}ipe {S}cience and {T}echnology}.
\newblock Global Digital Press, 1995.

\bibitem{reay2013}
D.~Reay, R.~McGlen, and P.~Kew.
\newblock {\em Heat {P}ipes: {T}heory, {D}esign and {A}pplications}.
\newblock Butterworth-Heinemann, 2013.

\bibitem{matsumoto1997}
S.~Matsumoto, S.~Maruyama, and H.~Saruwatari.
\newblock Molecular dynamics simulation of a liquid droplet on a solid surface.
\newblock {\em Japanese Journal of Tribology}, 42:143--152, 1997.

\bibitem{cao2006}
B.Y. Cao, M.~Chen, and Z.Y. Guo.
\newblock Liquid flow in surface-nanostructured channels studied by molecular
  dynamics simulation.
\newblock {\em Phys. Rev. E}, 74(6):066311, 2006.

\bibitem{nilson2006}
R.H. Nilson, S.W. Tchikanda, S.K. Griffiths, and M.J. Martinez.
\newblock Steady evaporating flow in rectangular microchannels.
\newblock {\em Int. J. Heat Mass Tran.}, 49(9-10):1603--1618, 2006.

\bibitem{yin2017}
H.~Yin, D.N. Sibley, U.~Thiele, and A.J. Archer.
\newblock Films, layers, and droplets: {T}he effect of near-wall fluid
  structure on spreading dynamics.
\newblock {\em Phys. Rev. E}, 95(2):023104, 2017.

\bibitem{holm1979}
F.W. Holm and S.P. Goplen.
\newblock Heat transfer in the meniscus thin-film region.
\newblock {\em J. Heat Transfer}, 101:543--547, 1979.

\bibitem{moosman1980}
S.~Moosman and G.M. Homsy.
\newblock Evaporating menisci of wetting fluids.
\newblock {\em J. Coll. Interf. Sci.}, 73:212--223, 1980.

\bibitem{do2008}
K.H. Do, S.J. Kim, and S.V. Garimella.
\newblock A mathematical model for analyzing the thermal characteristics of a
  flat micro heat pipe with a grooved wick.
\newblock {\em Int. J. Heat Mass Tran.}, 51:4637--4650, 2008.

\bibitem{akkus2016}
Y.~Akku{\c{s}} and Z.~Dursunkaya.
\newblock A new approach to thin film evaporation modeling.
\newblock {\em Int. J. Heat Mass Tran.}, 101:742--748, 2016.

\bibitem{akkus2017}
Y.~Akku{\c{s}}, H.I. Tarman, B.~{\c{C}}etin, and Z.~Dursunkaya.
\newblock Two-dimensional computational modeling of thin film evaporation.
\newblock {\em Int. J. Therm. Sci.}, 121:237--248, 2017.

\end{thebibliography}

\end{document}